\documentclass[fleqn,usenatbib]{mnras}

\usepackage{newtxtext,newtxmath}
\usepackage[utf8]{inputenc}
\hypersetup{draft}

\usepackage[T1]{fontenc}
\usepackage{ae,aecompl}

\usepackage{graphicx}	
\usepackage{amsmath}	
\usepackage{amssymb}	

\title[Origins of compact non-resonant systems]{The origins of nearly coplanar, non-resonant systems of close-in super-Earths}

\author[L. Esteves et al.]{
Leandro Esteves,$^{1}$\thanks{E-mail: leandro.esteves@unesp.br}
André Izidoro,$^{1,2}$\thanks{E-mail: izidoro.costa@gmail.com}
Sean N. Raymond,$^{3}$
Bertram Bitsch$^{4}$
\\
$^{1}$UNESP, Universidade Estadual Paulista, Grupo de Dinâmica Orbital e Planetologia, Guaratinguetá, CEP 12516-410, SP, Brazil\\
$^{2}$Department of Earth, Environmental and Planetary Sciences, MS 126, Rice
University, Houston, TX 77005, USA\\
$^{3}$Laboratoire d’astrophysique de Bordeaux, Univ. Bordeaux, CNRS, B18N, allée Geoffroy Saint-Hilaire, 33615 Pessac, France\\
$^{4}$Max-Planck-Institut für Astronomie, Königstuhl 17, 69117 Heidelberg, Germany\\
}

\date{Accepted XXX. Received YYY; in original form ZZZ}

\pubyear{2020}

\begin{document}
\label{firstpage}
\pagerange{\pageref{firstpage}--\pageref{lastpage}}
\maketitle

\begin{abstract}
Some systems of close-in ``super-Earths'' contain five or more planets on non-resonant but compact and nearly coplanar orbits. The Kepler-11 system is an iconic representative of this class of system. It is challenging to explain their origins given that planet-disk interactions are thought to be essential to maintain such a high degree of coplanarity, yet these same interactions invariably cause planets to migrate into chains of mean motion resonances. Here we mine a large dataset of dynamical simulations of super-Earth formation by migration. These simulations match the observed period ratio distribution as long as the vast majority of planet pairs in resonance become dynamically unstable. When instabilities take place resonances are broken during a late phase of giant impacts, and typical surviving systems have planet pairs with significant mutual orbital inclinations. However, a subset of our unstable simulations matches the Kepler-11 system in terms of coplanarity, compactness, planet-multiplicity and non-resonant state. This subset have dynamical instability phases typically much shorter than ordinary systems. Unstable systems may keep a high degree of coplanarity post-instability if planets collide at very low  orbital inclinations ($\lesssim1^\circ$) or if collisions promote efficient damping of orbital inclinations. If planetary scattering during the instability takes place at low orbital inclinations ($\text{i}\lesssim1^\circ$), orbital inclinations are barely increased by encounters before planets collide.When planetary scattering pumps orbital inclinations to higher values ($\gtrsim 1^\circ$) planets tend to collide at higher mutual orbital inclinations, but depending on the geometry of collisions mergers' orbital inclinations may be efficiently damped. Each of these formation pathways can produce analogues to the Kepler-11 system.

\end{abstract}




\section{Introduction}

At least 30\% of main sequence stars host planets with sizes between 1 and 4 Earth radii or masses less than 20~M$_{\earth}$ on orbital periods shorter than 100 days \citep{mayoretal11,howard12,winnfabrycky15}.  This population contains both high- and low-density planets~\citep{rogers15,wolfgang2016probabilistic}. Given that our focus is solely on their formation, we refer to this entire category of planets as {\em hot super-Earths}.  

Two key constraints on formation models are the observed multiplicity and period ratio distributions of super-Earths. Among transiting super-Earth systems only $\sim$20\% are found in multiple planets systems ({\rm N>2}). The remaining $\sim$80\% contain only a single {\em transiting} planet \citep{fang2012architecture,tremaine2012statistics,johansen2012can}. This has been called the Kepler dichotomy, and it is not clearly found in radial velocity systems~\citep[e.g.][]{figueiraetal12}. Adjacent super-Earths in multi-planet systems are rarely found in mean motion resonances \citep{fabrycky2014architecture}. Notable exceptions include the Kepler-223 \citep{millsetal16} and TRAPPIST-1 \citep{gillonetal17,lugeretal17} systems.

The migration model proposes that super-Earths formed at larger distances from their stars and migrated inwards by interactions with gaseous protoplanetary disks \citep{terquempapaloizou07,mcneilnelson10,idalin10}. In this model super-Earths pile up near the disk's inner edge, forming long chains of planets locked in first order mean motion resonances. After the gas disk dissipates (or just before), most resonant chains becomes dynamically unstable~\citep{izidoro2017breaking,izidoro2019formation}.  We call systems ``unstable'' if they undergo a phase of instability near the end of the gas disk lifetime and refer to the phase of subsequent collisions as the ``late instability phase''. Instability triggers scattering events among planets, breaks resonances and leads to giant impacts.  Close to the star collisions are favoured with respect to ejections because of high escape velocities from the star and short orbital periods \citep[e.g.][]{safronov72,cossouetal14,izidoro2017breaking,izidoro2019formation,raymondetal18,lambrechtsetal19,izidoro2019formation}. Surviving systems have orbits that are more spread out and dynamically excited than in the resonant chain phase. Systems that do not go dynamically unstable and stay in long resonant chains during the entire course of the simulations from \cite{izidoro2019formation} are refereed as ``stable'' systems. These simulations do no take into account other potential mechanisms that may help to trigger dynamical instabilities -- in systems that remained stable during the entire course of their simulations -- as for instance the effects of tidal dissipation, general relativity, planetesimal scattering and magnetospheric rebound \cite[e.g.][]{bolmont14,goldreichschlichting14,chatterjee2015planetesimal,liuetal17}. Here we assume that the instability phase is the source of the observed period ratio distribution of super-Earth systems.

Simulations show that the migration model matches the period ratio distribution of Kepler super-Earths if 90-99\% of resonant chains become unstable \citep{izidoro2017breaking,izidoro2019formation}. The same sample of systems matches the Kepler dichotomy because the considerable mutual inclinations between planetary orbits decreases the probability of finding many planets transiting in the same system \citep[see also][]{muldersetal18}. This argues that most single-transit systems are inherently multiple planet systems, in agreement with radial velocity surveys~\citep[e.g.][]{figueiraetal12}.

At first glance the migration model predicts that all stable systems should end up with many planets on coplanar, resonant orbits. When the viewing geometry is aligned with the planets' orbital plane, many planets should be detected in transit.  In contrast, unstable systems should produce planets on non-coplanar, non-resonant orbits. Rarely should many planets be found to transit in the same system in this case. 

There exists an observed category of system that is hard to understand in the framework of the migration model that contains many transiting planets on near-coplanar but non-resonant orbits. The Kepler-11~\citep{lissauer2011closely} and Kepler-20~\citep{fressin2012two,gautier2012kepler} systems are representative members of this class (see Fig. \ref{fig:fig1}). Six planets are known to transit in Kepler-11 with orbital periods between 10 and 118 days (semi-major axis between 0.091~au and 0.466~au)  and mutual orbital inclinations of at most 1 or 2 degrees~\citep{lissauer2013all}. The Kepler-20 system hosts two Earth-sized planets and four larger super-Earths, all with orbital periods between 3.69 and 77.61 days (semi-major axis between 0.045~au and $\sim 0.345 ~\text{au}$). Five of the six planets transit their star, but the second-outermost planet was discovered with the radial velocity technique~\citep{buchhave16}. Kepler-11's and Kepler-20's stellar ages have been estimated at 3.2$\pm$0.9 Gyr \citep{bedelletal17} and 7.6$\pm 3.7$ Gyr \citep{buchhave16}, respectively, indicating that each system is long-term stable~\citep[see also][]{migaszewskietal12,mahajanWu14,handsetal14,miakushvah16,jontofhutteretal17}.

Our paper is laid out as follows. In \S \ref{model} we present the setup of our simulations. Next we describe how we mined the outcome of these simulations to find 12 Kepler-11 analogue systems (\S \ref{goodsystems}). In \S \ref{formation} we show the typical formation pathway of these systems. We discuss our results and their implications in \S \ref{sec:discussion}.

\section{Simulations}\label{model}

We simulated the growth and dynamical evolution of systems of super-Earths during and after the gaseous disk phase~\citep[following from][]{izidoro2019formation}. Our code is based on the N-body code {\tt Mercury} \citep{chambers1999hybrid}, to which we have added synthetic forces designed to mimic essential planet formation processes. These include a prescription for disk evolution and dispersal~\citep[following][]{bitschetal15}, growth of planetary embryos by pebble accretion~\citep{johansenlambrechts17}, adopting a simple model for how the pebble flux evolves in time~\citep{izidoro2019formation}, tidal interactions between growing planets and the gas disk that lead to eccentricity and inclination damping~\citep{tanaka04,cresswell2008three} as well as orbital migration~\citep{ward97, paardekooper2011}. Collisions between growing bodies occur naturally and are treated as inelastic mergers, conserve mass and linear momentum. 

Our analysis is based on a subset of simulations from \citet{izidoro2019formation}, specifically their Model-I, II, and III runs. Our simulations start from a distribution of planetary seeds with masses of 0.005 to 0.015 Earth masses extending from either 0.7 to 20 AU (model I, \textit{a} and \textit{b} analogue indexes), 0.7 to 60 AU (model II, \textit{c}), or 0.2 to 2 AU (model III, \textit{d}).  Simulations of Model-I from \citet{izidoro2019formation} with nominal pebble flux and disk age (model I, \textit{b}) did not produce Kepler-11 analogues (see discussion below).

Seeds were given small but non-zero starting inclinations and eccentricities. Our simulations invoke a flux of pebbles spiralling inward through the disk due to aerodynamic drag~\citep[see, e.g., ][]{Lambrechts2014}. The inner edge of the gaseous disk was set at 0.1 AU, consistent with analyses of the Kepler super-Earths~\citep{muldersetal18} and as also found from radiation hydrodynamical simulations of the inner disc edge \citep{flocketal19}. The different models make different assumptions about the pebble size in the inner rocky parts of the disk as well as the disk mass at the start of the simulation, calibrated to match a given time in the evolution of our canonical disk model~\citep[taken from][]{bitschetal15}. In our simulations, seeds grow by accreting pebbles and eventually become massive enough to migrate (all the while continuing to accrete pebbles). Convergent migration leads to collisions among growing seeds, which feeds back on the migration rate. Growing planets may eventually reach pebble isolation mass \citep{bitschetal18}, where the pebble isolation mass in the inner disc regions could explain the presumed masses of the Kepler planets \citep[e.g.][]{wu2019,bitsch19}. After the dissipation of the gaseous disk, each simulation was integrated for another 50 Myr taking only gravitational perturbations into account. Simulations from \citet{izidoro2019formation} do not take into account gas accretion onto planetary embryos. For full technical details of the code and setup, see \citet{izidoro2019formation}, \citet{bitschetal19}, and \citet{lambrechtsetal19}.

In our analysis we made use of the subset of simulations from \citet{izidoro2019formation} that formed super-Earth systems roughly consistent with observations in terms of their masses~\citep[see][]{wolfgang2016probabilistic} and period ratio distributions~\citep[see][]{fabrycky2014architecture}.  This amounts to a total of 221 simulations.

\section{Mining our simulations for Kepler-11 and Kepler-20 analogues} \label{goodsystems}


Figure \ref{fig:fig1} shows the orbital architecture of twelve Kepler-11-like systems selected from our simulations.

We used four criteria to select Kepler-11 analogues, applied to planets within 0.7 AU: 
\begin{enumerate}
\item Systems could not contain more than one planet pair in first order mean motion resonance. The so called  stable systems do not match this first constraint so all our selected systems come from unstable systems of \cite{izidoro2019formation}.
\item Systems must have mean period ratios between adjacent planets within $\pm15\%$ of the Kepler-11 value of 1.68. 
\item Mutual inclinations between planetary orbits must be small enough that five or more planets could be observed in transit. 
\item Systems must maintain dynamical stability and criteria 1-3 over at least 1 billion years when we extend our simulations from 50~Myr to 1~Gyr.  
\end{enumerate}

\noindent{
\begin{figure}
	\includegraphics[width=0.4\paperwidth]{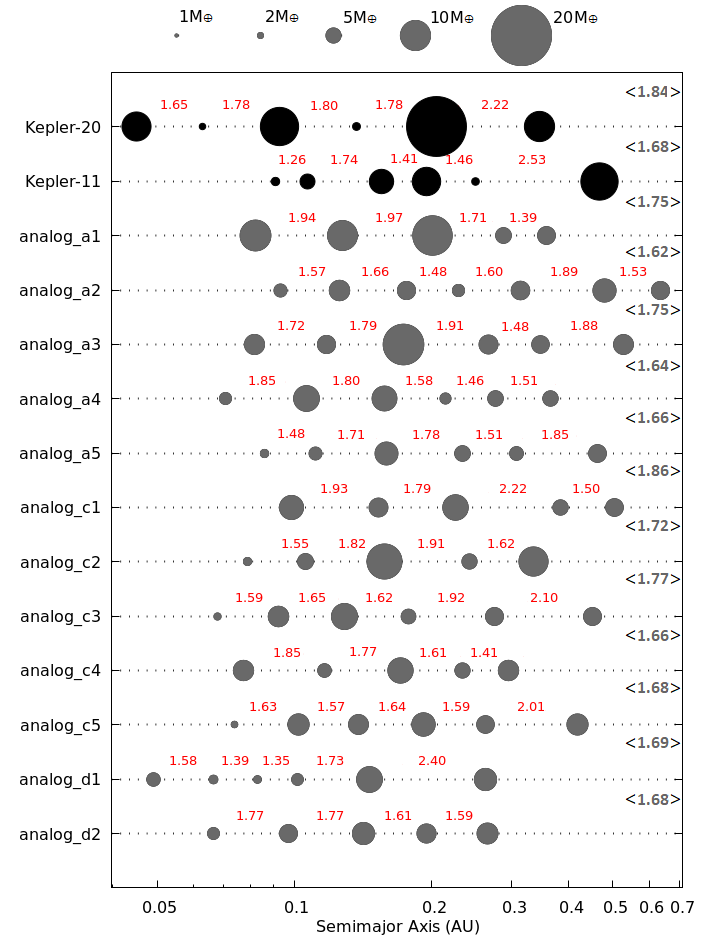}
    \caption{A selection of simulated planetary systems with dynamical architectures similar to those of Kepler-11 and Kepler-20 (shown at the top). Each horizontal line shows a full system. The planet size scales with mass. For the estimated masses of Kepler-11 and Kepler-20 planets we follow \citet{bedelletal17} and \citet{buchhave16}, respectively. As estimated masses are not available for Kepler-11g, Kepler-20e, and Kepler-20f we estimated their masses using the mass-radius relationship from \citet{wolfgang2016probabilistic}. The red numbers between planet pairs show the period ratio of the respective planet pair. The black numbers on the far right show the mean period ratio of each system. Size of the dots scales linearly with mass as shown at the top. The planets in the figure have orbital eccentricities ranging from 0.0034 to 0.099, and the whole simulated sample mean eccentricity is 0.0292. We recall that simulations from \protect\citet{izidoro2019formation} do not take into account gas accretion onto planetary embryos.}
    \label{fig:fig1}
\end{figure}
}

We will discuss how the fraction of Kepler-11 like systems in our simulations compare to observations in Section \ref{sec:discussion}.


 A nice example simulation -- \textit{analog\_c5} -- contains 6 planets between 0.07 AU and 0.42 AU. Only the outermost pair of super-Earths appears to be in 2:1 resonance (in fact one of the resonant angles associated with the 2:1 mean motion resonance librates and circulates). The mean period ratio between planets in this system is 1.688, very close to the Kepler-11 value of 1.68 (The mean period ratio of the Kepler-20 planet pairs is $\sim$1.84). Considering the estimated masses of the Kepler-11 and Kepler-20 planets \citep{lissauer2011closely,bedelletal17,buchhave16,wolfgang2016probabilistic} we have also calculated the  mean mutual Hill radii of these systems. To calculate the mean Hill radius of each system we first calculated the 
 mutual Hill radii of adjacent-planet pairs in the system and then we average over the pairs in the system.
 
 Using the upper and lower limits of their estimated masses, the mean mutual hill radii of the Kepler-11 system are 12.85 and 14.45, respectively. For the Kepler-20 system, these metrics ranges between 14.35 and 16.10. The mean separation in mutual Hill radii of our twelve selected systems are between 13 and 16.5. The mean mutual Hill radii of \textit{analog\_c5} is 14.76. This shows that the compactness of our selected systems compares well with those of Kepler-11 and Kepler-20 systems not only in terms of orbital period ratio of adjacent planet pairs but also in terms of their mutual Hill radius. Thus, both Kepler-11 and Kepler-20 systems are consistent as being the outcome of dynamical instabilities rather than being necessarily the outcome of type-I migration in inviscid disks that may produce compact systems that fail to form resonant chains \citep{mcnally2019multiplanet}.

\noindent{
\begin{figure}
	\includegraphics[width=0.42\paperwidth]{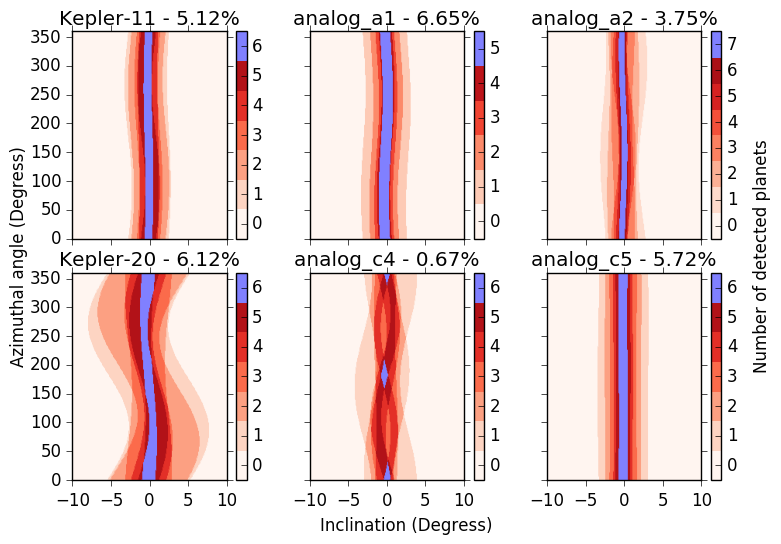}
    \caption{A geometric transit map for a model Kepler-11, Kepler-20 and four simulated systems \protect\citep[data from][respectively]{bedelletal17,buchhave16}. The vertical axis represents the azimuthal line of sight whereas the horizontal axis is the inclination of an observer's line of sight relative to the system's ecliptic plane. The color scale indicates the number of planets that transit for each observing geometry. The percentage next to each system name indicates the fraction of viewing angles from which all planets transit. Orbital inclination data for planet Kepler-20g and eccentricity of Kepler-20d are uncertain, so we assumed the values 0.2 and 1.3 degrees, respectively, within data upper limit.}
    \label{fig:panel}
\end{figure}
}

To estimate the probability of finding multiple planet systems, we simulated the geometric transit probability from a range of observing angles.  Figure~\ref{fig:panel} shows how the detectability of five simulated systems varies as a function of the azimuthal angle $\phi_{\text{obs}}$ of an observer along a reference plane and the inclination of the observer i$_{\text{obs}}$ with respect to that same plane. Each axis was divided into 200 points and at each point and for each planet in a given system, we calculated the height of the planet relative to the system reference plane z$_{\text{planet}}$ and the height of the observer z$_{\text{obs}}$ relative to same plane. We imposed a minimum impact parameter of 0.9 (such that $\text{z}_{\text{det}} \lid 0.9 \text{r}_{\text{star}}$) to indicate a transit detection for a given planet.

The color scale in Fig.~\ref{fig:panel} illustrates the number of planets detected in transit at each line of sight. For a almost perfectly coplanar system the region the number of planets seen in transit is roughly independent of the azimuthal angle and varies simply with the observer's inclination relative to the plane of the planets. That would amount to a vertical line in Fig.~\ref{fig:panel}.  However, in some systems (e.g., \textit{analog\_c4}) certain azimuthal angles are preferred, representing special alignments, e.g., where the longitudes of ascending node of multiple planets cross. The amplitude of inclination that each planet covers in the figure is inversely proportional to its distance from the star.


Fig.~\ref{fig:panel} shows that in many simulated systems five or more planets could be detected in transit. Of course, this only represents a snapshot in time. To understand the long-term evolution of these systems, and to ensure that their transit detection is maintained, we integrated these systems for 1 Gyr past their final configuration at 50 Myr obtained from \citet{izidoro2019formation}. We only included planets within 0.7 AU to reduce the computing time need. 

The systems in Fig.~\ref{fig:fig1} are those that maintained dynamical stability during their long-term integrations. We have extended the simulations of \cite{izidoro2019formation} using the hybrid integration algorithm of Mercury package \citep{chambers1999hybrid}. We use an integration timestep smaller than 1/20 of the orbital period of innermost planet in the system.
Any additional instability phase would reduce the system planet multiplicity, spread out the planets' orbits, and increase mutual inclinations, likely making it impossible to detect many planets in transit. Five of our initial candidates in fact became unstable in our long-term integrations and were discarded, and they are not shown in Figure \ref{fig:fig1}. Fig.~\ref{fig:long} shows the evolution of the planets' inclinations in four systems that remained stable for 1 Gyr. The color scale represents the number of planets seen in transit for an observer at $\phi_{\text{obs}} = 0$ and i$_{\text{obs}} = 0$ over the systems evolution. Two systems (analogues a1 and c5) maintained extremely low mutual inclinations and the full 5-6 planet systems are seen in transit throughout. This may be analogous to Kepler-11 (see Fig.~\ref{fig:panel}).  However, in two other cases (a4 and c3) the number of planets seen in transit changes in time as the planets exchange orbital angular momentum, inducing fluctuations in mutual inclinations. This may be representative of the Kepler-20 system, in which only 5 of 6 known planets transit. 

\noindent{
\begin{figure}
	\includegraphics[width=0.42\paperwidth]{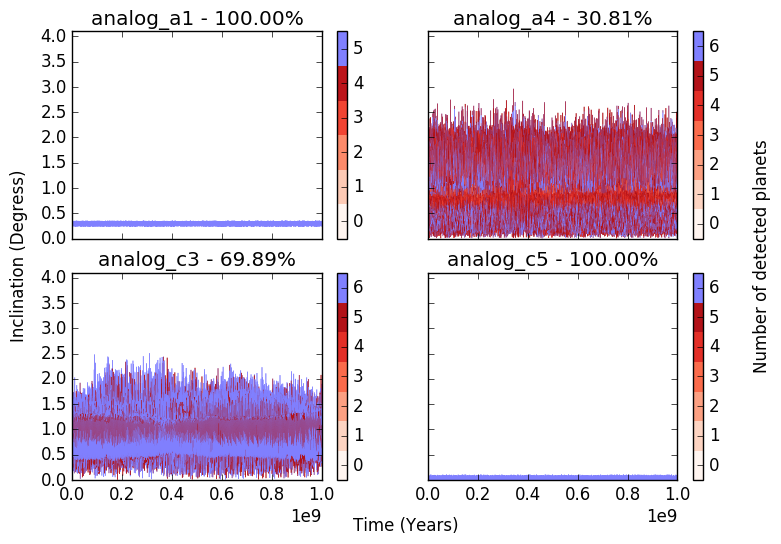}
    \caption{Long-term evolution of the planets' orbital inclination in four different systems simulated for 1 billion years. At each time interval, the colors indicate the number of planets in transit. Each panel show the orbital inclinations of all planets in the system. Upper-left and bottom-right panels show that planets' orbital inclinations in these systems vary over a very small range. The percentage next to the name of each system represents the fraction of the integration during which all of the planets in a given system would be found to transit.}
    \label{fig:long}
\end{figure}
}

\section{Formation pathways of Kepler-11 analogue systems}\label{formation}

We now investigate the formation of the Kepler-11 analogue systems selected in the previous section.  The top panel of Figure \ref{fig:damping} shows the mean orbital inclination of planets just before they collide (when they touch) during the instability phase and the final mergers' orbital inclinations. Each point/star represents the mean value calculated over all planets colliding in a given system during the instability phase. Kepler-11 like systems come in two flavours. They have either  very low orbital inclinations when they collide or exhibit very efficient damping of orbital inclination by collisions. Kepler-11 like systems have final planets with mean orbital inclinations typically lower than $\sim$1 degree. It is possible to understand these results  from the lower panel of the same figure.

 While the top panel of Figure \ref{fig:damping}  show systems' mean values, its lower panel shows the orbital inclination of individual planets (in fact this is the largest inclination of any two colliding bodies) just before they collide during the instability phase. The vertical axis show mergers inclinations just after the collisions. Again, stars show Kepler-11 like systems, but now  small-dots show  systems with at least 5 planets that do not qualify as Kepler-11 like systems following our criteria. The color shows the difference of the longitude of the ascending nodes of  colliding planet pairs just before the collision. 

  When planetary scattering during the instability phase takes place very close to the system's invariant plane ($i<h_{\text{rmh}}=R_{\text{hill}}/a = \frac{(m_i+m_j)}{3M_{\text{star}}} \lesssim 1^\circ$; $h_{\text{rmh}}$ is the reduced mutual Hill radius)  the orbital inclination increases very slowly by encounters \citep{ida1990} and planets created by mergers tend to have very low inclination orbits, independent on the bodies orbital alignment during the collision \citep[][see orange data points in the lower panel of Figure \ref{fig:damping}]{matsumotoetal17}.  If the orbital inclination of colliding bodies is sufficiently low ($i \lesssim i_{\text{esc}}=0.5v_{\text{esc}}/v_k$; where $v_{\text{esc}}$ and $v_k$ are the  escape (combining the masses of $i$ and $j$) and Kepler velocity, respectively) the $|v_z|$ velocity component is small and $v_z$ is accelerated due to the gravitational focusing during the approach \citep{matsumotoetal17}.  If the velocity change is larger than $iv_k$ then the longitude of the ascending nodes of the approaching bodies change such as ${\Omega_i - \Omega{_j} \simeq 180^\circ}$ and inclination gained during the encounter is mostly damped by the collision  \citep[][see black data points in the lower panel of \ref{fig:damping}]{matsumotoetal17}. If bodies eventually reach $i \gtrsim i_{\text{esc}}=0.5v_{\text{esc}}/v_k$ during planetary scattering, gravitational focusing may  not  have time to align the nodes before the impact \citep{matsumotoetal17}. In this case, the damping of orbital inclination may not be very efficient (e.g. red and green dots in the lower panel of Figure \ref{fig:damping}). Damping of orbital inclination occurs in most collisions (Figure \ref{fig:damping}) in agreement with \citet{matsumotoetal17}. Collisional damping of orbital inclination may occur at any distance from the star. However, because collisions are more likely to occur close to the star due to shorter dynamical timescales and higher escape velocities this inclination damping mechanism is probably far more efficient in close-in regions than further out.

\noindent{
\begin{figure}
	\includegraphics[width=0.41\paperwidth]{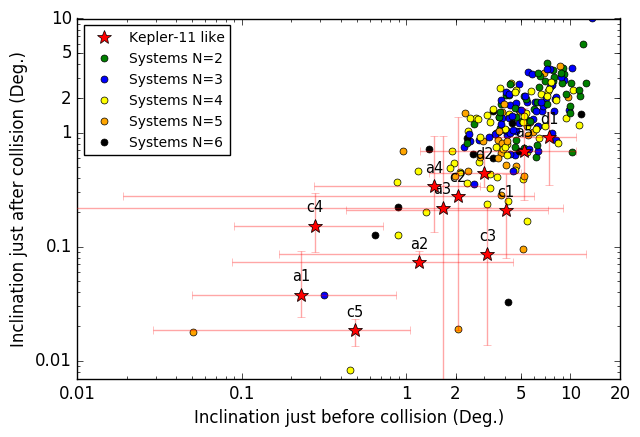}
     \includegraphics[width=0.41\paperwidth]{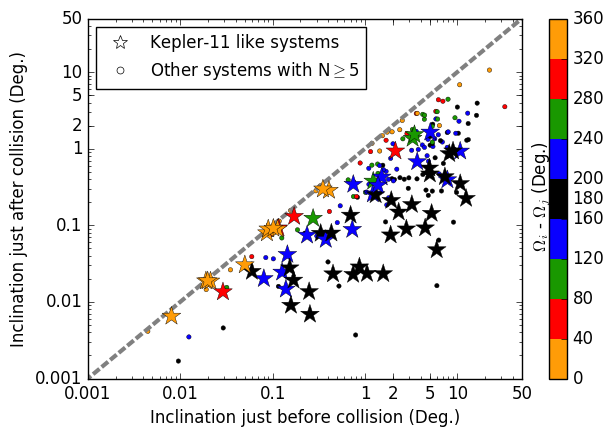} 
     \caption{{\bf Top panel:}  Mean orbital inclination of planets just before collisions in the instability phase and the final mergers' mean orbital inclinations. Each data point shows the mean values of one planetary system.  The mean orbital inclination just before collision of each system is calculated averaging over all colliding planet pairs of the system but taking the maximum inclination of each planet-pair. Kepler-11 like systems are shown as red stars. Non Kepler-11 like systems appear as dots where dot-colors represent the number of planets in the system inside 0.7~AU. Inclinations are given with respect to gas disk midplane. The vertical and horizontal error bars show lower and upper bounds of the sample over which the respective average were calculated. Labels near the stars show the analogue name as also shown  in Figure \ref{fig:panel}. {\bf Bottom panel:} Inclination of the merged planet as a function of the largest inclination of any two colliding bodies just before the collision. Each data point represent a single collision, unlike the top panel that shows the mean value of all collisions. Color-coding is used to show the alignment of the longitude of the ascending node of the colliding bodies. The dashed gray line shows the identity function. Data points vertically distant from the identity show collisions that dissipated significant inclination while those near the identity function barely affected orbital inclinations.  }
    \label{fig:damping}
\end{figure}
}

Figure~\ref{fig:k11_1} shows the growth of the planets in our best Kepler-11 analogue, the \textit{analog\_c5} system. Shortly before the final dissipation of the gas disk a dynamical instability was triggered ( see for instance \cite{goldreichschlichting14} and \cite{pichierrietal18,pichierietal20} for a detailed discussion on the onset of dynamical instabilities in resonant chains). While it led to a number of collisions,  merger planets had very low (mutual) orbital inclinations because planetary scattering during the instability phase took place near the system invariant plane (in this case is also near the disk midplane) such that orbital inclinations grew very slowly \citep{ida1990}. As collisions conserve linear momentum, the merger planets also have very low mutual orbital inclinations. The instability phase in this system was quite short and ended before the gas fully dissipated. After the instability phase, there was insufficient time for planets to restore a resonant configuration before the gas fully vanished. In fact planets did not migrate significantly post-instability in this simulation. Although it is difficult to infer the exact role of the gas tidal effects for the final outcome of the dynamical instability (but see also \cite{kominamiida02} and \cite{iwasakietal01}) the presence of the gas-disk during the stability phase is not always a required condition to produce our Kepler-11 like systems. This is the case for example for \textit{analog\_a1}. In this system, the instability phase started after the gas dispersal yet the planets remained on extremely low-inclination orbits ($\sim0.1^\circ$). This inclination range is comparable to that of planets in \textit{analog\_c5} (Figure \ref{fig:k11_1}). For  the \textit{analog\_a1}, the key ingredient to produce a almost coplanar system is the fact that planet collided at super-low mutual orbital inclinations \citep{matsumotoetal17}. For systems like \textit{analog\_a1}, the eventual anti-alignment of the longitude of the ascending nodes at collisions is a mere bonus that help to damp already low orbital inclinations even further. About half of our Kepler-11 analogues underwent instabilities after the gas disk was fully gone.




\noindent{
\begin{figure*}
	\includegraphics[width=0.8\paperwidth]{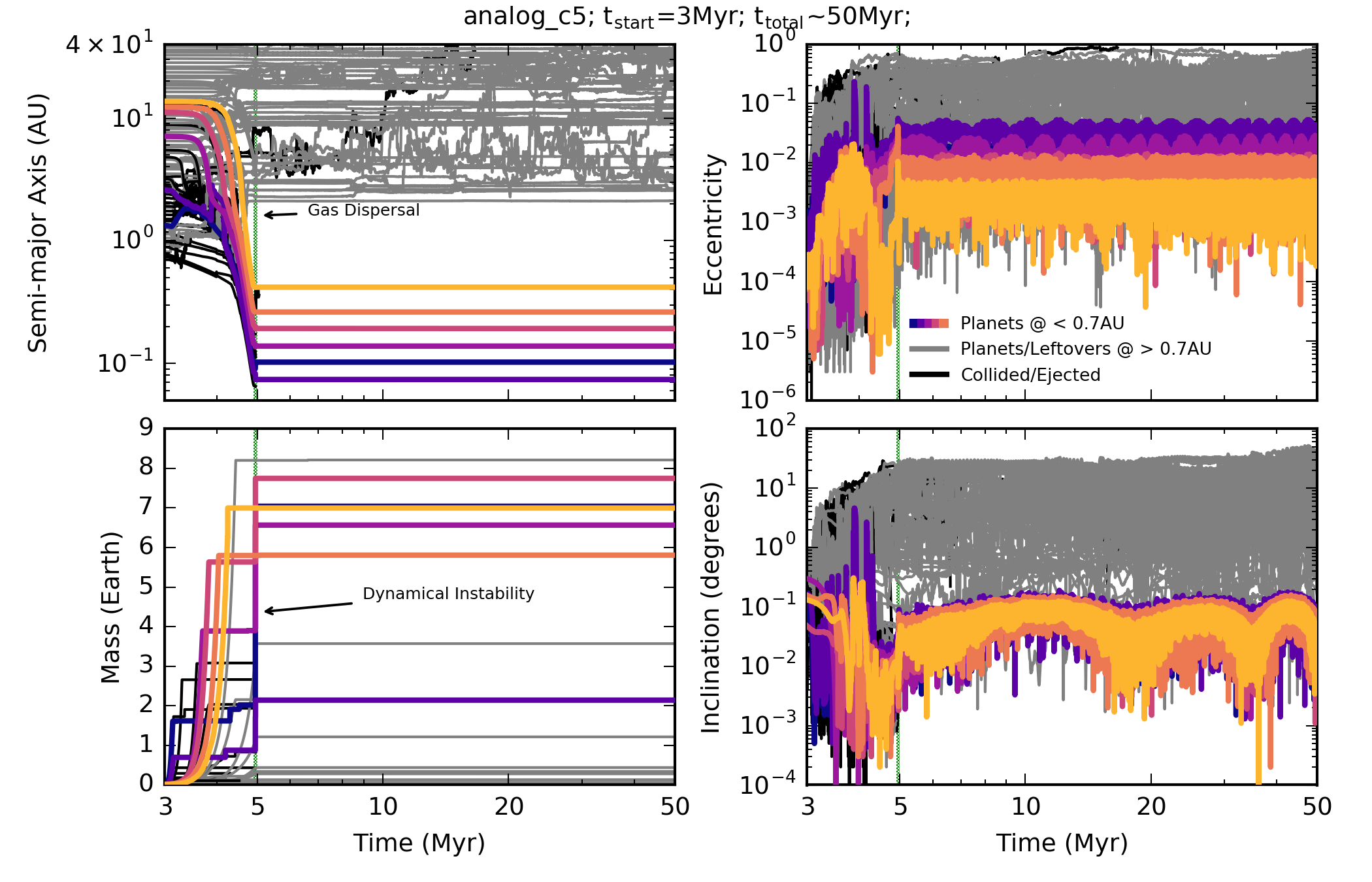}
    \caption{Formation of the \textit{analog\_c6} system.  The different panels show the evolution of the growing planets' orbital radii (top left), eccentricity (top right), mass (bottom left) and inclination (bottom right). Only planets within 0.7 AU are shown in color. More distant objects are shown in gray and those that underwent collisions in black. This simulation started at $\text{t}_{\text{start}} = 3$~Myr. The green line shows when the last phase of gas disk dissipation (from 4.9 to 5 Myr), during which a strong but short-lived instability led to a number of collisions.}
    \label{fig:k11_1}
\end{figure*}
}

The shortest instability phases of our simulated Kepler-11 like systems were for planets that collided at low orbital inclination (e.g. \textit{analog\_a4}, \textit{analog\_c5}, and \textit{analog\_a2}).  We define the duration of the instability phase as the $\Delta_{t}$ between the timing of the first and last collision in the system.  We only take into account collisions occurring later than 100~kyr before gas disk dispersal and inside 0.7~au. We expect the instability duration to depend on how fast resonances between planet pairs are broken, the orbital inclinations/eccentricities of planet pairs when this happens, and the number of planet pairs \citep[see][]{pichierrietal18}. Unstable pairs on almost coplanar orbits collide almost immediately while inclined ones typically take longer. As the number of planets and consequently potential collisions in different unstable resonant chains may vary, we normalize the duration of the instability phase of each system ($\Delta_{t}$)  by the respective number of collisions during the instability phase ($\Delta_t / N_{\text{col}}$). The quantity $\Delta_t / N_{\text{col}}$  measures the averaged time between successive collisions during the instability phase, and it is a good proxy to measure how fast collisions take place in different resonant chains. We find that in our Kepler-11 like systems the timing between successive collision is almost always shorter than 0.1~Myr (about 86\% of successive collisions). 
For a  2-$\sigma$ confidence level  $\Delta_t / N_{\text{col}}$ yields (0.278 $\pm$ 0.433)~Myr  in  Kepler-11 analogues. For non Kepler-11 analogues (with N > 4) it yields: (2.39 $\pm$ 1.51)~Myr. In Kepler-11 like systems, collisions happen fast and presumably before inclinations are significantly excited by encounters.

We also did look for other correlations that could exist between the timing of the instability phase and the dynamical state of the systems but we did not find any obvious one. This is true for instance when we measure the onset of the dynamical instability relative to the end of the gas disk dispersal (we did check this correlation for other convenient epochs). The timing of onset of the dynamical instability probably depends on  the complex resonant dynamics of each resonant chain. On the other hand, the duration of the impact phase -- measured via $\Delta_t / N_{\text{col}}$ -- may provide insights on the dynamical state of the system when resonances are broken.
 
In our simulations collisions are considered to be perfectly inelastic and conserve mass and linear momentum. To validate this assumption we have confronted the impact data of our simulations with merging criteria for giant impacts. We find that most giant impacts during the instability phase in our simulations qualify as merging events \citep[e.g.][]{gendakokuboida12,kokubogenda10,leinhardt2012,stewart12}. Figure \ref{fig:impacts} shows the normalized impact velocities of collisions during the late instability phase with colors corresponding to merging (black) and hit-and-run (red) collisions. Only 10\% of collisions in our Kepler-11 systems fall in the hit-and-run regime; the other 90\% are merging. Thus, on average, less than one collision in Kepler-11 systems is a hit-and-run. Among our full sample of simulations the fraction of hit-and-run collisions is about 31\%. Our simplistic treatment of giant impacts is also supported by the results of \citet{poon19}, who found virtually no difference between planetary systems where collisions are considered perfect merging events and where fragmentation is taken into account~\citep[see also][]{wallace17}. The effects of fragmentation also have minor effects on the formation of terrestrial planets in our solar system \citep{chambers13,clement19b}.

\noindent{
\begin{figure}
	\includegraphics[width=0.41\paperwidth]{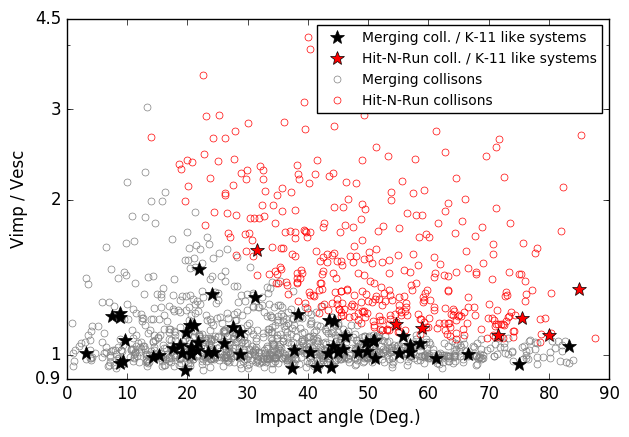}
    \caption{ Normalized impact velocities for giant impacts from our simulations as a function of the impact angle. Only collisions occurring during the instability phase are shown. In addition, only are shown collisions where the bodies involved have masses larger than 0.6 Earth mass. Collisions in Kepler-11 like systems are shown as stars and non-selected systems are shown as circles. In both cases, the black (circles or stars) show collisions that qualify as merging events and the red (circles or stars) show collisions that would qualify as hit-and-run collisions. The outcome of our collisions are classified following the criteria by \protect\citet{kokubogenda10} and \protect\citet{gendakokuboida12}. }
    \label{fig:impacts}
\end{figure}
}

Finally, in Kepler-11 analogues systems the planets had also lower orbital eccentricities immediately before collisions when compared with most typical simulations, this can be seen clearly in Figure \ref{fig:ea_vs_eb}. For a 2-$\sigma$ confidence level the mean orbital eccentricity of planet-pairs just before collisions is (0.176 $\pm$ 0.0176) for Kepler-11 like systems, while for non Kepler-11 like systems it is (0.253 $\pm$ 0.00762). This also supports our claim that as in Kepler-11 like systems collisions typically happen very rapidly there is virtually no time for planets to dynamically over excite each other's orbit by mutual scattering and encounters. Figure \ref{fig:ea_vs_eb}  also shows that collisions can efficiently damp orbital eccentricities which in agreement with previous studies \citep[e.g.][]{raymondetal06,matsumotoetal15}.


\noindent{
\begin{figure}
	\includegraphics[width=0.41\paperwidth]{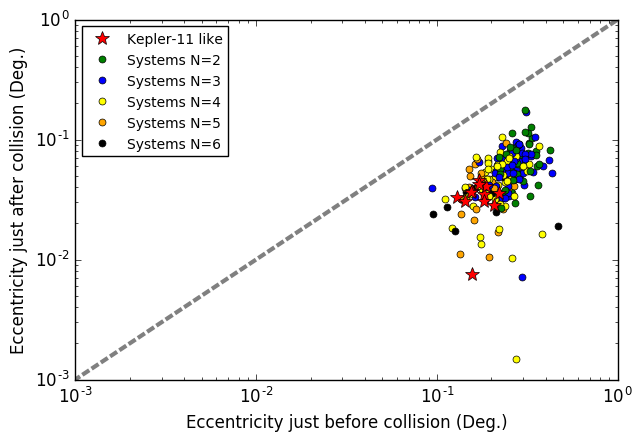}
    \caption{ Mean orbital eccentricity of planets just before collisions in the instability phase and the final mergers’ mean orbital eccentricities. Each data point shows one planetary system. The mean orbital eccentricity just before collision of each system is calculated averaging over all colliding planet pairs of the system but taking the maximum eccentricity of each planet-pair. Kepler-11 like systems are shown as red stars. Non Kepler-11 like systems appear as dots where dot-colors represent the number of planets in the system inside 0.7AU. }
    \label{fig:ea_vs_eb}
\end{figure}
}



\section{Discussion} \label{sec:discussion}


We have shown that Kepler-11-like systems, with highly coplanar but non-resonant orbits, are a natural outcome of the migration model. Of course, they only represent a small sub-population of hot super-Earths. Yet their formation pathway -- characterized by collisions at low orbital inclinations or collisions that efficiently damp orbital inclinations due to the anti-alignment of the longitude of the ascending node of colliding bodies -- falls within the range of plausible outcomes of our simulations. Such systems are therefore not extreme rarities but simply modestly-rare outcomes of a more general formation scenario characterized by migration of planets into resonant chains, followed by late instabilities~\citep[the so-called {\em breaking the chains} model;][]{izidoro2017breaking,izidoro2019formation}. 

One may wonder whether long-range migration is an essential ingredient in producing Kepler-11 analogue systems. We do not think that this is the case.  Many different models have been proposed to explain the population of close-in super-Earths~\citep[see][for a comparison between models]{raymond08}. A common theme in all models is that super-Earths must form quickly, likely reaching near their final masses during the gaseous disk phase. This implies that the late stages of the gaseous disk phase are likely characterized by the migration of super-Earths, regardless of where and how they actually accreted~\citep{raymond18}. The {\em breaking the chains} model proposes that, like the awkward teenage years, resonant chains are a phase through which all super-Earth systems pass, although migration need not always imply resonant chains~\citep{mcnally2019multiplanet}. Yet the formation of Kepler-11 analogues relies on the geometry of impacts during the instability phase.


The fraction of Kepler-11 like systems varies between different sets of simulations from \citet{izidoro2019formation}. Their Model I -- with their nominal pebble fluxes and disk ages -- produced planets that were typically more than 10 Earth masses. Systems with such massive planets are typically more excited and dynamically spread and none of these systems qualified as a Kepler-11 like system considering our criteria.  Kepler-11 like configurations should be more common in lower-mass systems. \citet{izidoro2019formation}'s Model-II produced five Kepler-11 like systems, with an occurrence rate of about 10\% of the simulations of this model. To compare our results with observations, we conducted simulated observations of our planetary systems and we find that Model-II systems produced 6.4 times more systems with 3 or more observed planets than systems with 5 or more planets. For the Kepler data, this same ratio is much higher, about 17.  So, it is quite possible that some of our model overestimates the fraction of high  multiplicity systems. However, it is important to keep in mind that high planet multiplicity systems comes in two flavours in the context of the breaking the chain scenario: from stable resonant chains such as Kepler-223 or from unstable Kepler-11 like systems. Matching the Kepler data may also require a combination of different disk models \citep{izidoro2019formation}. Thus, we caution when comparing the fraction of Kepler-11 like systems produced in our simulations and those in the Kepler data.

Our simulations are admittedly simplified and we do not claim them to be the final answer in planet formation.  Nonetheless, they include a diversity of relevant physical effects thought to represent many of the key processes in planet formation.  It is also reassuring that the systems we produce bear a strong resemblance to observed systems~\citep[see][for a discussion]{izidoro2019formation}.

We did not obtain systems identical to Kepler-11 and Kepler-20. Systems with very compact orbits, such as the first pair of the Kepler-11 system, almost always became unstable in our simulations. However, we note that migration can produce pairs of planets on very close orbits~\citep{raymondetal18}, which may in principle be gently separated on Gyr timescales through tidal friction~\citep{lithwick12,batygin13,bolmont14}.   

Looking to the future, it would be interesting to take the compositions and atmospheric envelopes of super-Earths into account together with a more sophisticated model for giant impacts, to see whether there is a correlation with coplanar non-resonant systems. It may also be important to include fragmentation in giant impact models as debris produced in the dynamical instabilities can break resonant chains if the set of debris is massive enough \citep{chatterjee2015planetesimal}. Many planets in the Kepler-11 and Kepler-20 systems have low densities \citep{lissauer2013all} although Kepler-20b may have a terrestrial composition \citep{fressin2012two,gautier2012kepler}.

\section*{Acknowledgements}

 We thank the referee, Ramon Brasser, for his very constructive comments and suggestions that help us to improve the paper. L. E. and A. I. are grateful to FAPESP for financial support through grants 17/09963-7 and 19/02936-0 (L. E.), 16/12686-2 and 16/19556-7 (A. I.). B.B. thanks the European Research Council (ERC Starting Grant 757448-PAMDORA) for their financial support. A. I. acknowledges NASA grant 80NSSC18K0828 to Rajdeep
Dasgupta, during the  revision and ressubmission of the work. \\




\bibliographystyle{mnras}
\bibliography{references} 



\appendix

\section{Secular resonances in our systems}
 In order to further understand the dynamics of our planetary systems, we have also mined our data looking for main secular resonances in our final planetary systems.  From simulated planet-pairs in Figure \ref{fig:fig1}, 7 planet-pairs exhibit significant (and in most cases multiple) episodes of libration of the resonance angle associated with the  main inclination-type secular resonance, while the resonant angles associated with the main eccentricity-type secular resonance mostly circulate (Figure \ref{fig:ires}). Additionally, we found another 5 planet-pairs where the resonant angle associated with the main secular eccentricity-type resonance showed significant (and in most cases multiple) episodes of libration, while the resonant angles associated with the main inclination-type secular resonance mostly circulated  (e.g. Figure \ref{fig:eres}, left panel). Only 1 of our planet-pairs show episodes of libration of both angles associated with the eccentricity and inclination types main secular resonances. We consider significant episodes of libration those that last at least 1-10Myr.
 
\noindent{
\begin{figure}
	\includegraphics[width=0.41\paperwidth]{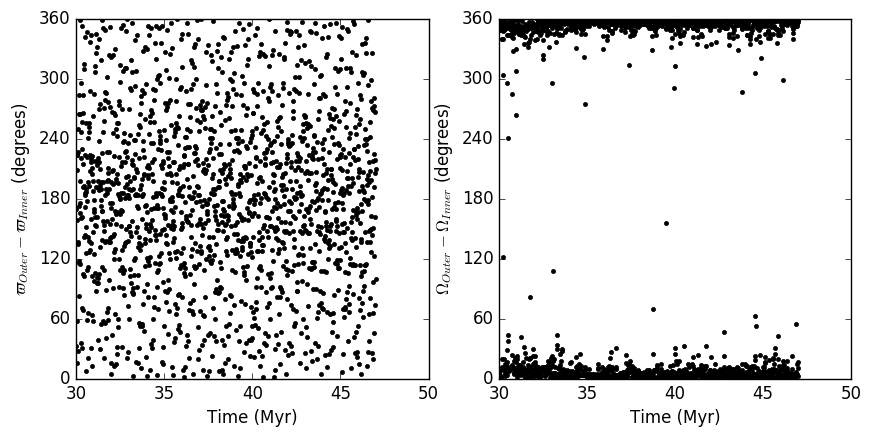}
    \caption{ Time evolution of the resonance angle  associated with the  eccentricity-type  main secular resonance (left-panel). As shown this angle is circulating. On the other hand, the right panel shows that the resonant angle associated with the inclination-type main secular resonance is mostly librating around 0$^{\circ}$, so this planet-pair is near/in an inclination-type secular resonance. These angles are computed considering the second and third innermost planets in the \textit{analog\_a1} system.}
    \label{fig:ires}
\end{figure}
}

\noindent{
\begin{figure}
	\includegraphics[width=0.41\paperwidth]{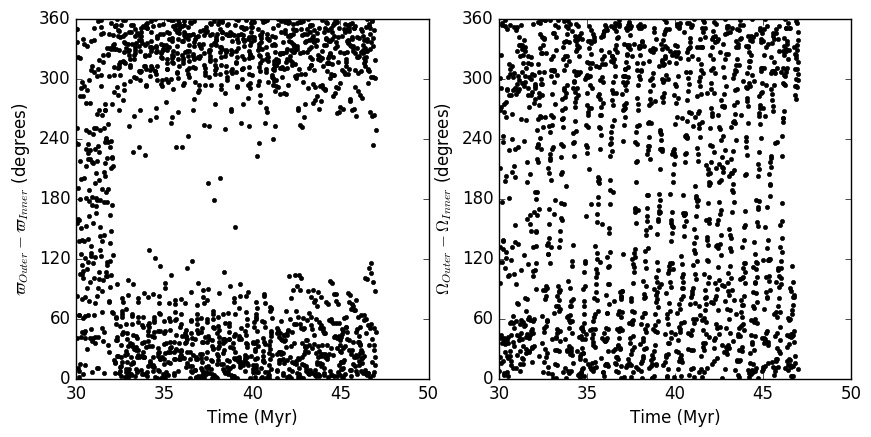}
    \caption{ Same panels as in figure \ref{fig:ires}. Here the resonant angle associated with the eccentricity-type main secular resonance  shows episodes of libration with large amplitude around 0$^{\circ}$, while the resonant angle associated with the inclination-type main secular resonance mostly circulates.  These resonant angles have been computed considering the two outermost planets of \textit{analog\_c3} system.}
    \label{fig:eres}
\end{figure}
}



\bsp	
\label{lastpage}
\end{document}